\documentclass[%
 aip,
rsi,%
 amsmath,amssymb,
 reprint,%
]{revtex4-1}

\usepackage[dvipsnames]{xcolor}
\usepackage{graphicx}% Include figure files
\usepackage{dcolumn}% Align table columns on decimal point
\usepackage{bm}

\usepackage{dcolumn}% Align table columns on decimal point
\usepackage{epstopdf}
\usepackage{xcolor}
\usepackage{subcaption} 
\usepackage{float}
\usepackage{tabularx}
\usepackage{mathrsfs}
\usepackage{upgreek}% non-italic greek symbols
\usepackage[a]{esvect}
\setcitestyle{square}
\usepackage[toc,page]{appendix}
\usepackage[font={small,normalfont}]{caption}
\usepackage[colorlinks=true, allcolors=blue]{hyperref}

\begin{document}

\title{Energy gain by laser-accelerated electrons in a strong magnetic field}

\author{A. Arefiev}
\affiliation{Department of Mechanical and Aerospace Engineering, University of California at San Diego, La Jolla, CA 92093}

\affiliation{Center for Energy Research, University of California at San Diego, La Jolla, CA 92093}

\author{Z. Gong}
\affiliation{SKLNPT, School of Physics, Peking University, Beijing 100871, China}
\affiliation{Center for High Energy Density Science, The University of Texas, Austin, TX 78712}

\author{A. P. L. Robinson}
\affiliation{Central Laser Facility, STFC Rutherford-Appleton Laboratory, Didcot, OX11 0QX, UK}

\date{\today}

\begin{abstract}
The manuscript deals with electron acceleration by a laser pulse in a plasma with a static uniform magnetic field $B_*$. The laser pulse propagates perpendicular to the magnetic field lines with the polarization chosen such that $(\bm{E}_{laser} \cdot \bm{B}_*) = 0$. The focus of the work is on the electrons with an appreciable initial transverse momentum that are unable to gain significant energy from the laser in the absence of the magnetic field due to strong dephasing. It is shown that the magnetic field can initiate an energy increase by rotating such an electron, so that its momentum becomes directed forward. The energy gain continues well beyond this turning point where the dephasing drops to a very small value. In contrast to the case of purely vacuum acceleration, the electron experiences a rapid energy increases with the analytically derived maximum energy gain dependent on the strength of the magnetic field and the phase velocity of the wave. The energy enhancement by the magnetic field can be useful at high laser amplitudes, $a_0 \gg 1$, where the acceleration similar to that in the vacuum is unable to produce energetic electrons over just tens of microns. A strong magnetic field helps leverage an increase in $a_0$ without a significant increase in the interaction length.
\end{abstract}

\maketitle

%*******************************************************

\section{Introduction} \label{Sec-intro}

Direct laser acceleration (DLA) is a robust mechanism for generating large populations of energetic electrons in laser-irradiated plasmas~\cite{pukhov1999DLA,Mangles-PhysRevLett.94.245001,Nilson_2010,arefiev2016beyond}. It is also a reliable way to transfer the energy of an irradiating laser pulse to the plasma. One advantage of the direct laser acceleration is that it generates forward-directed electrons at relativistic laser intensities. The energetic electrons can then be leveraged to produce secondary particle (ion~\cite{Daido_2012,Macchi_2013_RMP}, neutron~\cite{higginson2011production,Pomerantz-PhysRevLett.113.184801}, positron\cite{cowan_positron,Chen_2009_PRL,Chen_2010_PRL}) and radiation sources~\cite{Schwoerer_ph,huang2016characteristics,Stark2016PRL}. 

A useful reference point for the performance of DLA is the energy gain by an initially immobile electron irradiated by a plane electromagnetic wave in a vacuum~\cite{gibbon2004short,kruer2018physics}. The maximum energy that the electron can achieve in a wave with a normalized amplitude $a_0$ is
\begin{equation} \label{DLA_vac}
    \varepsilon_0 = \gamma_0 m_e c^2 = \left( 1 + \frac{a_0^2}{2} \right) m_e c^2,
\end{equation}
where $a_0$ is defined in terms of the wave electric field $E_0$ and frequency $\omega$ as
\begin{equation}
    a_0 \equiv \frac{|e| E_0}{m_e c \omega}.
\end{equation}
Here $m_e$ and $e$ are the electron mass and charge and $c$ is the speed of light. One can also relate the normalized amplitude to the wave intensity $I_0$ and wavelength $\lambda$, with $a_0 \approx 0.85  \sqrt{I_0 [10^{18} \mbox{ W/cm}^{2}]} \lambda[\mu \mbox{m}]$. Equation~(\ref{DLA_vac}) indicates that an electron can achieve $\varepsilon_0 \approx 20$~MeV in a laser pulse with $I_0 \approx 10^{20}$~W/cm$^{2}$ and $\lambda = 1$~$\mu$m ($a_0 \approx 8.5$).

One difficulty of extrapolating this result to higher $I_0$ is that the acceleration distance increases with laser intensity. The electron has to travel a considerable distance with the laser pulse before it is able to achieve the energy given by Eq.~(\ref{DLA_vac}). The corresponding distance roughly scales as $\Delta \propto \gamma_0 \lambda \propto a_0^2 \lambda$. At $a_0 \approx 50$, we have $\Delta > 100 \lambda$. In the case of a tightly focused laser pulse, this $\Delta$ is much longer than the Rayleigh length. 

Experiments aimed at measuring DLA in a plasma have shown that the electron energy can exceed $\gamma_0 m_e c^2$~(see Refs.~[\onlinecite{gahn1999PRL_DLA,Mangles-PhysRevLett.94.245001}]). %but that the energy gain can occur over a longitudinal distance that is much shorter than $\gamma_0 \lambda$. 
The departure from the purely vacuum acceleration regime has been attributed to the presence of quasi-static electric fields that arise in a plasma when the interaction exceeds the characteristic electron response time~\cite{pukhov1999DLA,willingale_channeling,Arefiev_2012_PRL,arefiev2016beyond}. Even though these fields are much weaker than the field of the laser $\max (E_{wave})$, they profoundly alter the electron dynamics to enhance $-(\bm{v} \cdot \bm{E}_{wave})$ or/and to prolong the time when $-(\bm{v} \cdot \bm{E}_{wave}) > 0$~(for example, see Refs.~[\onlinecite{arefiev2016beyond}] and [\onlinecite{Khudik-POP_2016}]. The increased work by the laser on the electron then leads to an improved energy gain compared to the purely vacuum regime.

In contrast to the static electric fields, the role of quasi-static magnetic fields has remained relatively unexplored in the context of the direct laser acceleration. At currently achievable laser intensities, the magnetic field generated in the plasma is typically weaker than the plasma electric field~(see examples provided in Refs.~[\onlinecite{arefiev_JPP_2015}] and [\onlinecite{arefiev2016beyond}]), so it is then not surprising that the laser-driven magnetic field is only of secondary importance in this regime. The next generation of laser facilities is projected to reliably achieve on-target intensities exceeding $10^{22}$~W/cm$^{2}$~[\onlinecite{ELI,XCELS,APOLLON.2017}]. Numerical simulations performed in anticipation of achieving these intensities have shown that such an intense laser pulse is able propagate through a classically overdense plasma and drive a very strong longitudinal plasma current ($\sim$MA)~\cite{gong2018_FSSA}. This current can then generate and sustain a strong magnetic field $(\sim$MT)~\cite{ji2014radiation,Stark2016PRL,Jansen_2018}. The simulations have also revealed that the plasma electric fields in overdense plasmas become suppressed due to a reduced ion response time~\cite{Jansen_2018,gong2018_FSSA}. The combination of the reduction in the electric field and the increase in the magnetic field means that a regime with a dominant quasi-static magnetic field will become accessible in overcritical plasmas irradiated by high-intensity laser pulses. 

In order to gain better insight into direct laser acceleration in the presence of quasi-static magnetic fields, we consider a simplified setup where an electron is irradiated by a plane electromagnetic wave in a uniform magnetic field that is transverse to the laser propagation. We are specifically interested in determining how much energy an electron with an initial transverse momentum can gain over a time interval of just a single cyclotron period. This setup captures a key element that would be also relevant to electron dynamics in the nonuniform laser-driven magnetic field. 

The rest of the manuscript consists of five sections. Section~\ref{Sec-basic} provides the basic equations that describe the electron dynamics in our setup. An example of the direct laser acceleration in a uniform magnetic field is given in Sec.~\ref{Sec-DLA_with_B}. Section~\ref{Sec-estimates} gives estimates for the maximum attainable energy and the corresponding spatial displacement. Detailed parameter scans obtained by numerically solving the equations from Sec.~\ref{Sec-basic} and confirming the robustness of the estimates are given in Sec.~\ref{Sec-parameter_scan}. Section~\ref{Sec-superluminosity} examines the impact of the superluminossity on the direct laser acceleration process. The results are summarized in Sec.~\ref{Sec-summary}, where we also provide additional comments to emphasize the importance of the obtained results.

%*******************************************************

\section{Basic equations} \label{Sec-basic}

The dynamics of a relativistic electron is described by the following equations:
\begin{eqnarray}
&& \frac{d \bm{p}}{d t} = - |e| \bm{E} - \frac{|e|}{\gamma m_e c} \left[ \bm{p} \times \bm{B} \right], \label{EQ_1} \\
&& \frac{d \bm{r}}{d t} = \frac{c}{\gamma} \frac{\bm{p}}{m_e c}, \label{EQ_2}
\end{eqnarray}
where $\bm{r}$, $\bm{p}$ are the electron position and momentum, $t$ is the time, 
\begin{equation}
    \gamma = \sqrt{1 + p^2 /m_e^2 c^2}
\end{equation}
is the relativistic factor, and $\bm{E}$ and $\bm{B}$ are the electric and magnetic fields acting on the electron. In the regime under consideration, $\bm{E} = \bm{E}_{wave}$ is just the laser electric field, whereas $\bm{B} = \bm{B}_{wave} + \bm{B}_*$ is a superposition of the magnetic field of the laser and the static uniform magnetic field $B_*$. 

In order to simplify our analysis, we approximate the laser pulse as a plane linearly polarized electromagnetic wave with a given phase velocity $v_{ph}$. Without any loss of generality, we assume that the wave propagates along the $x$-axis and we set
\begin{eqnarray}
&& \bm{E}_{wave} = \bm{e}_y E_0 \cos \left(s + 2 \pi \psi \right), \label{E_wave} \\
&& \bm{B}_{wave} = \bm{e}_z \frac{c}{v_{ph}} E_0 \cos \left(s + 2 \pi \psi \right), \label{B_wave}
\end{eqnarray}
where $E_0$ is the wave amplitude, $\psi$ is the phase offset,
\begin{equation} \label{s}
s \equiv \omega t - \omega x / v_{ph}
\end{equation}
is the phase of the wave with frequency $\omega$ at the electron's location. 

We consider a configuration where the uniform magnetic field is directed along the $z$-axis and the electron has no momentum along the magnetic field lines, so that the electron trajectory remains flat. It is then convenient to introduce the following notations:
\begin{equation}
    \bm{p} = \bm{e}_x p \cos \theta + \bm{e}_y p \sin \theta,
\end{equation}
where $p$ is the absolute value of the momentum and $\theta$ is the angle between the momentum vector and the direction of the laser propagation.

The two non-trivial components of Eq.~(\ref{EQ_1}) can be arranged as equations for $\theta$ and $\gamma$:
\begin{eqnarray}
    && p \frac{d \theta}{d t} = -|e| E_y \cos \theta + |e| B_z \frac{v}{c} , \\
    && \frac{d \gamma}{d t} = -|e| E_y \sin \theta \frac{p}{\gamma m_e^2 c^2}.
\end{eqnarray}
After taking into account the considered field configuration we find that
\begin{eqnarray}
    \frac{d \theta}{d ( t \omega )} & = & -a_0 \cos \left(s + 2 \pi \psi \right) \frac{1}{\gamma}  \frac{c}{v}  \left[ \cos \theta - \frac{v}{v_{ph}} \right] \nonumber \\ &+& \frac{1}{\gamma}  \frac{\omega_{ce}}{\omega}, \label{main_1} \\
    \frac{d \gamma}{d ( t \omega )} & = & - \frac{a_0 p}{\gamma m_e c} \sin \theta \cos \left(s + 2 \pi \psi \right), \label{main_2}
\end{eqnarray}
where
\begin{equation}
    a_0 \equiv \frac{|e| E_0}{m_e c \omega}
\end{equation}
is the dimensionless laser amplitude and 
\begin{equation}
    \omega_{ce} = \frac{|e| B_*}{m_e c}
\end{equation}
is the non-relativistic electron cyclotron frequency. 

%*******************************************************

\section{Example of DLA in a magnetic field} \label{Sec-DLA_with_B}

\begin{figure}
    \begin{center}
\includegraphics[width=1\columnwidth]{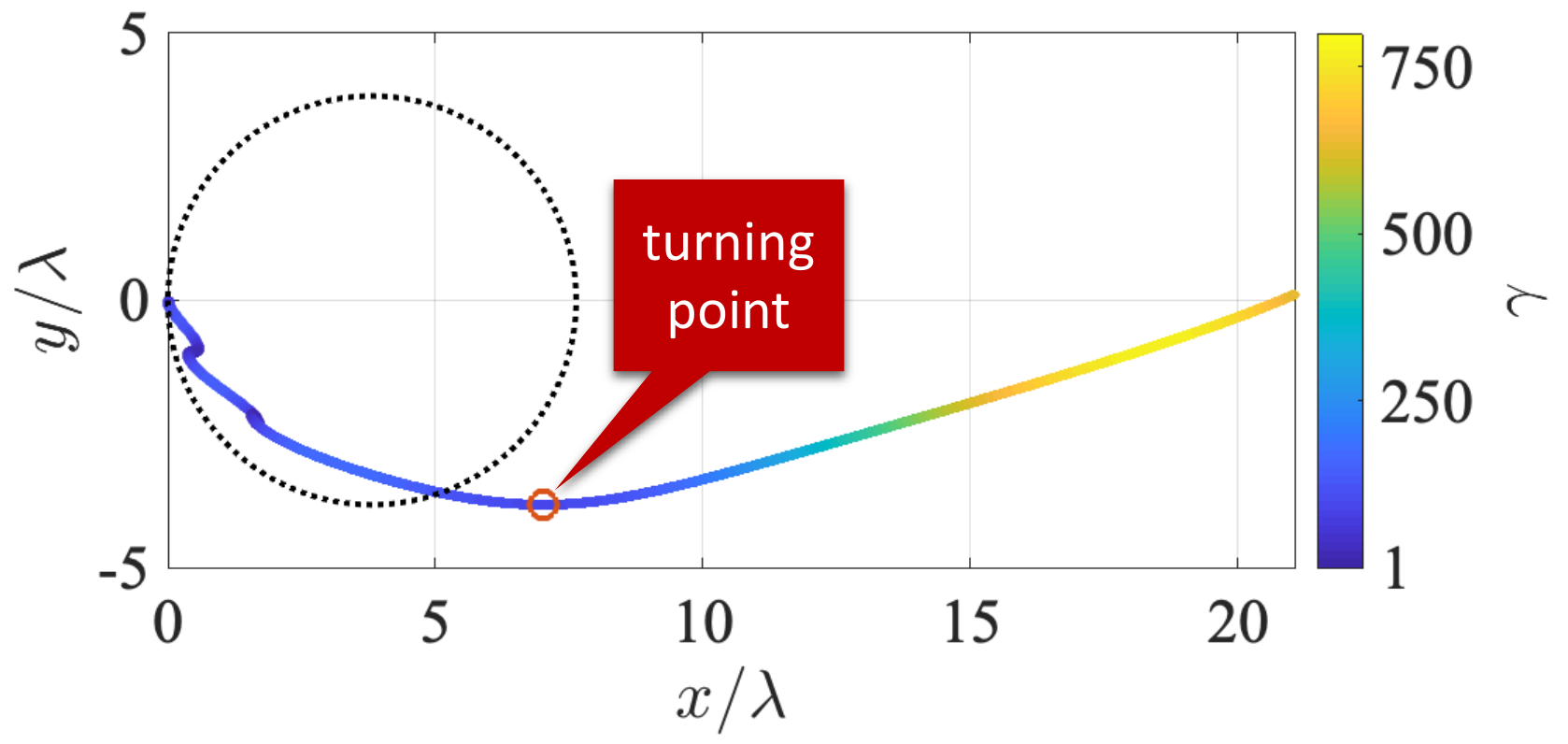}
       \caption{\label{fig:traj} Electron trajectories in a uniform magnetic field with $\omega_{ce}/\omega = 2.085$. The solid line is for an electron with an initial transverse momentum $p_y = -50 m_ec$ that is irradiated by a laser pulse with $a_0 = 50$. The dotted line is for the same electron but without the laser pulse.}
       \end{center}
\end{figure}

In order to examine the effect of a strong magnetic field on DLA, we consider an electron that starts its motion in the laser pulse with a transverse momentum $\bm{p}_0 = (0, -p_0,0)$ at $s = 0$. There is no phase offset in this case, $\psi = 0$, so the electron starts its motion in the strongest laser field. We also set $v_{ph} = 0$ in this example to make an easier connection with the published results for purely vacuum DLA without an additional static magnetic field.

In the case without the magnetic field, the solution is well known:
\begin{equation} \label{gamma_DLA_vac}
    \gamma = \frac{1}{2R} \left[ 1 + R^2 + \left( a_0 \sin s + p_0 / m_e c \right)^2 \right],
\end{equation}
where
\begin{equation}
    R = \frac{\gamma}{\omega} \frac{ds}{dt}
\end{equation}
is the so-called dephasing rate. The dephasing rate is a constant of motion in a plane wave and it is equal to $R = \gamma - p_x / m_e c$. We take into account that $\bm{p}_0 = (0, -p_0,0)$ at $s = 0$ to find that $R = \sqrt{1 + (p_0 / m_e c)^2}$.

We consider an example with $p_0 / m_e c = a_0 \gg 1$. We then have $R \approx a_0$. According to Eq.~(\ref{gamma_DLA_vac}), the electron reaches its maximum energy at $s = \pi / 2$, with
\begin{equation} \label{EQ_18}
    \max (\gamma) \approx 5 a_0 /2 \ll \gamma_0 \approx a_0^2 /2. 
\end{equation}
The transverse motion is clearly detrimental, because the maximum $\gamma$-factor is less than the maximum $\gamma$-factor for an initially immobile electron, given by Eq.~(\ref{DLA_vac}). The underlying cause is a high dephasing rate that decreases the time the electron spends gaining the energy from the laser electric field before it slips into a decelerating phase.

\begin{figure}
    \begin{center} 
    \includegraphics[width=1\columnwidth]{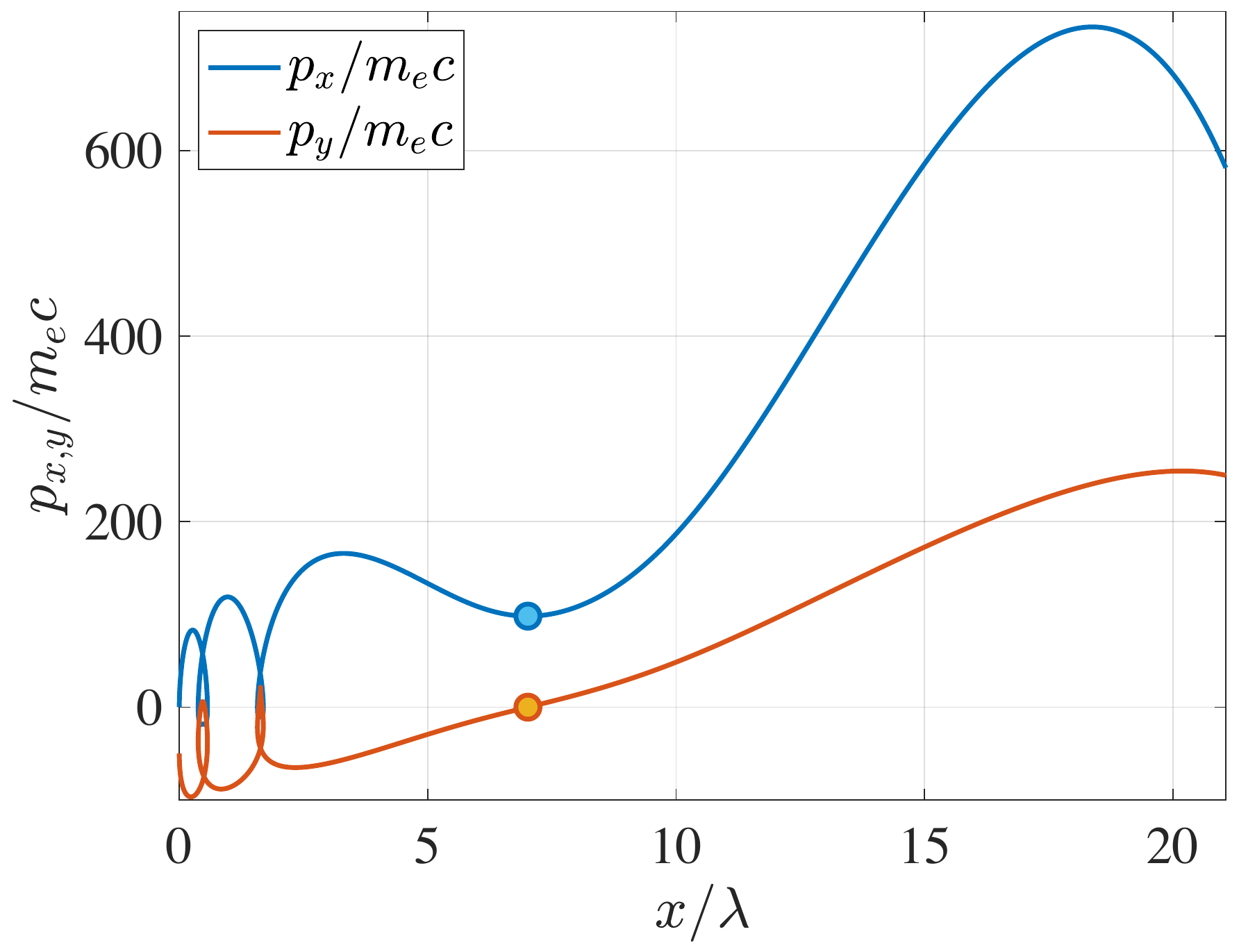}
       \caption{\label{fig:traj_p} Longitudinal and transverse components of the electron momentum along the color-coded trajectory shown in Fig.~\ref{fig:traj}. The circles mark the turning point from Fig.~\ref{fig:traj}.}
       \end{center}
\end{figure}

As seen in Fig.~\ref{fig:traj}, a strong static magnetic field with $\omega_{ce}/\omega = 2.085$ dramatically enhances the energy gain of an electron in a plane wave with $a_0 = 50$. The initial conditions are the same as in the previous example, but the maximum relativistic factor is now $\max (\gamma) \approx 770$, which is at least six times higher than $\max (\gamma)$ without the magnetic field [see Eq.~(\ref{EQ_18})].

\begin{figure}
    \begin{center}
\includegraphics[width=1\columnwidth]{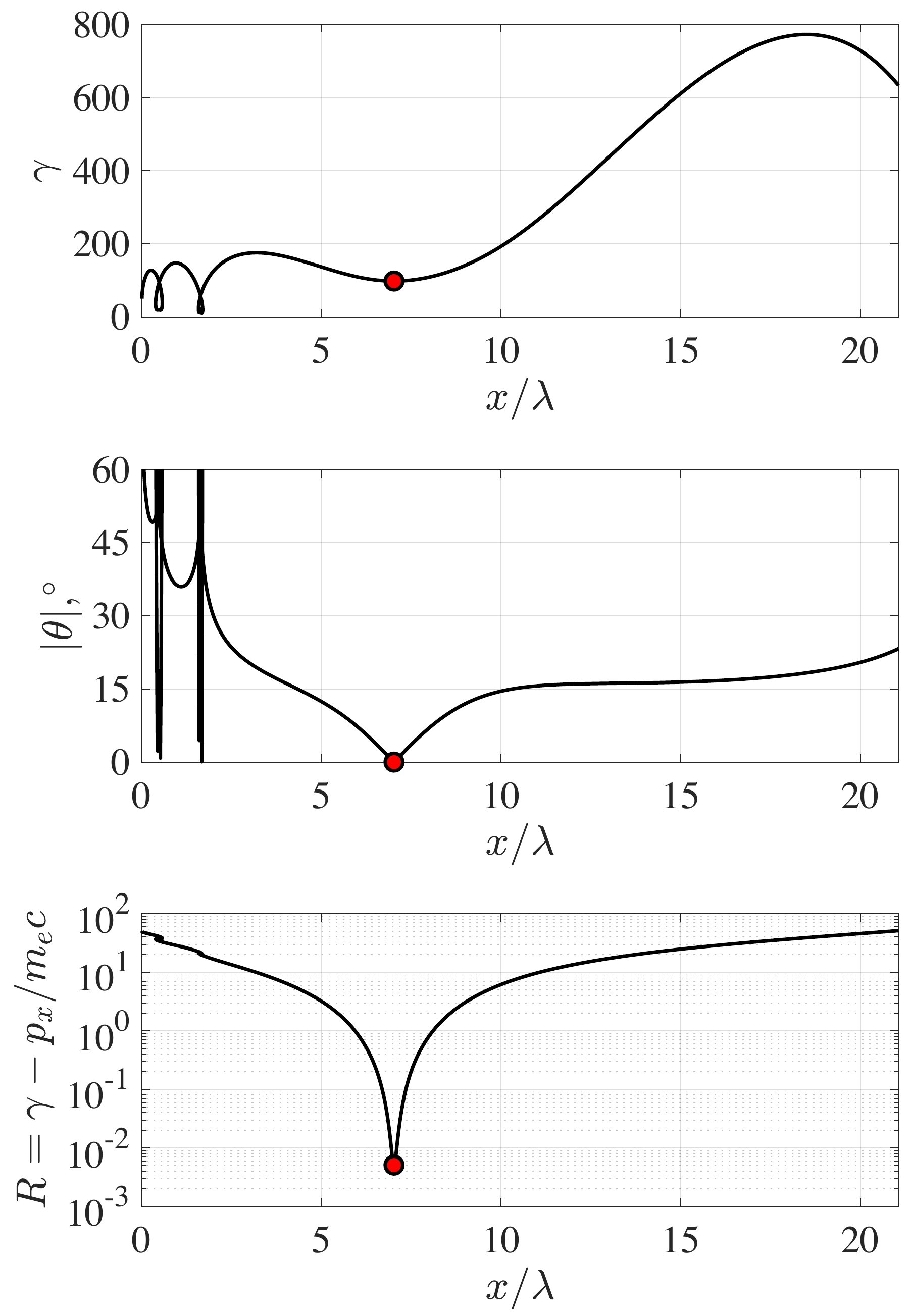}
       \caption{\label{fig:2} Relativistic factor $\gamma$, angle $\theta$, and the dephasing rate $R$ of the laser-irradiated electron in a uniform magnetic field. The circles mark the turning point from Fig.~\ref{fig:traj}.}
       \end{center}
\end{figure}

The key difference is the rotation of the momentum by the static magnetic field that reduces the dephasing between the electron and the wave. The dotted trajectory in Fig.~\ref{fig:traj} is the gyro-orbit of the electron in the absence of the laser field. The corresponding rotation period is
\begin{equation}
    T = \frac{2 \pi}{\omega_{ce}} \sqrt{1 + \frac{p_0^2}{m_e^2 c^2}}.
\end{equation}
After a quarter of this period, the momentum is pointing forward and the dephasing rate formally calculated using the expression
\begin{equation} \label{R}
    R = \gamma - p_x / m_e c
\end{equation}
yields $R \approx m_e c / 2 p_0 \ll 1 $. As seen in Fig.~\ref{fig:2}, the dephasing calculated along the trajectory of the laser-irradiated electron (Fig.~\ref{fig:traj}) confirms the same trend: the dephasing gradually reduces as the electron approaches the bottom of its trajectory where the momentum is directed forward (see Fig.~\ref{fig:traj_p}).

Even though the enhanced energy gain is triggered by the dramatic reduction in the dephasing by the magnetic field, most of the energy gain occurs at relatively high values of $R$, with $R \gg 1$. Indeed, the energy gain in Fig.~\ref{fig:2} takes place as the electron moves at an angle of roughly 16$^\circ$ to the $x$-axis. According to Eq.~(\ref{R}), we have $R \approx p \theta^2 /2$, where $p$ is the total momentum. As $p$ increases, so does the dephasing $R$ (instead of remaining at a constant low value). Therefore, the acceleration in the presence of the magnetic field qualitatively differs from the conventional vacuum acceleration with low initial dephasing that remains constant.

We have confirmed using different values of the magnetic field, electron momentum, and $a_0$ that the enhancement always occurs after the electron passes the bottom part of its trajectory, i.e. the turning point. There is one consistent feature: the energy enhancement starts at $E_{wave} < 0$ when the electron momentum is directed forward, with $p_x / m_e c \gg 0$. To make this point more evident, the circles in Figs.~\ref{fig:traj_p} and \ref{fig:2} mark the values at the turning point of the trajectory shown in Fig.~\ref{fig:traj}. 

%*******************************************************

\section{Estimates for DLA in a magnetic field} \label{Sec-estimates}

In what follows, we perform simple estimates to identify the key features of the direct laser acceleration in a uniform magnetic field. The estimates are based on trends discussed in Sec.~\ref{Sec-DLA_with_B}.

In order to provide the context for our estimates, we first review the main features of the direct laser acceleration in a vacuum. We consider an electron that starts its motion from rest at the moment when $E_y = - E_0$. This corresponds to $\psi = -1/2$ and $p_x = p_y = 0$ at $s = 0$. The solution for the electron's momentum in a laser pulse with $v_{ph} = c$ is
\begin{eqnarray}
    && p_x /m_e c = \frac{1}{2} a_0^2 \sin^2 (s), \\
    && p_y /m_e c = a_0 \sin (s).
\end{eqnarray}
We are interested in a high-amplitude laser pulse with $a_0 \gg 1$ that can accelerate electrons to ultra-relativistic energies. As the electron accelerates and its momentum becomes relativistic, the angle $\theta$ decreases. We find directly from the provided solution that for $\gamma \gg 1$ we have
\begin{equation} \label{theta_DLA}
    \theta_{vac} \approx \sqrt{2/\gamma}.
\end{equation}

One of the main weaknesses of the direct laser acceleration in vacuum is that the energy transfer from the laser to the electron becomes inefficient with the energy increase. This point is evident from Eq.~(\ref{main_2}) where the rate of the energy increase is proportional to $\sin \theta \approx \theta$. We have shown that $\theta \propto \gamma^{-1/2}$, which indicates that the rate of the energy transfer becomes suppressed as $\gamma^{-1/2}$. The suppression reflects the fact that the electron moves almost forward, so that its velocity is nearly orthogonal to the electric field of the laser that does the work on the electron. A direct consequence of this is that the electron has to travel a significant distance with the laser pulse in order to reach its maximum energy, with $\gamma_{\max} = 1 + a_0^2/2$, for $a_0 \gg 1$. For example, this distance is roughly $\Delta x \approx 150 \lambda$ for $a_0 = 50$ , where $\lambda$ is the laser wavelength. 

A static magnetic field alters the energy exchange with the laser by preventing the angle $\theta$ from decreasing with the energy increase. The last term in Eq.~(\ref{main_1}) for $d \theta / dt$ counterbalances the first term that causes the already discussed reduction in $\theta$. Our goal is to find the corresponding angle $\theta$ where the reduction stops. The corresponding condition reads
\begin{equation} \label{EQ-balance}
    -a_0 \cos \left(s + 2 \pi \psi \right) \frac{1}{\beta}  \left[ \cos \theta - \frac{\beta}{u} \right] +  \frac{\omega_{ce}}{\omega} = 0,
\end{equation}
where
\begin{eqnarray}
    && \beta \equiv v/c, \\
    && u \equiv v_{ph}/c.
\end{eqnarray}
It is evident from the structure of this equation that the smallest value of $\theta$ allowed by the magnetic field corresponds to the strongest laser field, with $ -a_0 \cos \left(s + 2 \pi \psi \right) \approx a_0$. Using this approximation and by taking into account that the angle is small, we find that
\begin{equation} \label{EQ-balance2}
   \frac{\theta^2}{2} \approx \frac{u - \beta}{u} +  \frac{\beta}{a_0} \frac{\omega_{ce}}{\omega}.
\end{equation}

Equation~(\ref{EQ-balance2}) provides a general scaling for the angle between the electron momentum and the $x$-axis, so it it instructive to consider limiting cases. In the limit of $u \rightarrow 1$ and $\omega_{ce} \rightarrow 0$, we have $\theta^2 \approx 2(1-\beta)$. In this case, $\theta \ll \theta_{vac}$, where $\theta_{vac}$ is the smallest angle achieved during the purely vacuum acceleration and it is given by Eq.~(\ref{theta_DLA}). This means that the considered compensation never occurs in this regime. If the superluminosity is important but the magnetic field is still weak, we have $\theta^2/2 \approx (u - \beta)/u$ where we need to set $\beta \approx 1$. As a result we find that the angle is given by
\begin{equation}
    \theta_{ph} = \sqrt{2(u-1)} = \sqrt{2 \delta u},
\end{equation}
where
\begin{equation}
    \delta u = u - 1 = (v_{ph} - c)/c
\end{equation}
is the measure of the superluminosity. This result matches the result that we previously derived in Ref.~[\onlinecite{Robinson_PoP_2015}]. If the magnetic field dominates the acceleration process, then the last term in Eq.~(\ref{EQ-balance2}) dominates. We set $\beta \approx 1$ to find that the corresponding angle is given by
\begin{equation} \label{theta_min}
    \theta_{mag} = \left(\frac{2\omega_{ce}}{a_0 \omega} \right)^{1/2}.
\end{equation}

\begin{figure}
    \begin{center}
\includegraphics[width=1\columnwidth]{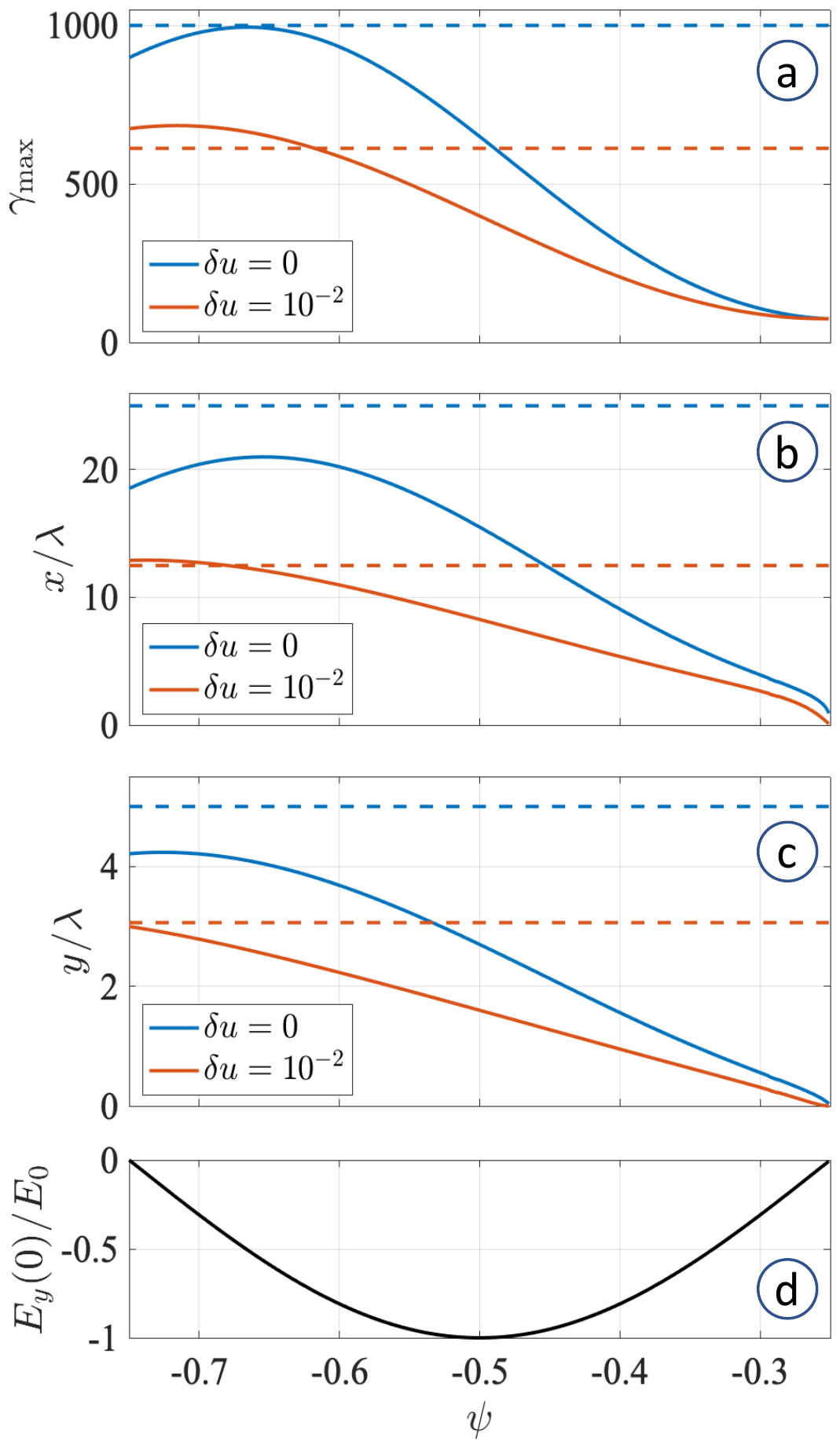}
       \caption{\label{fig:phase_scan} Scan over the phase offset for an electron with $\bm{p}_0 = (75 m_e c, 0 ,0)$ irradiated by a wave with $a_0 = 50$ in a magnetic field with $\omega_{ce}/ \omega = 1$. (d) the initial amplitude of the laser electric field. (a) - (c) the maximum relativistic factor $\gamma_{\max}$ and the maximum displacement that the electron achieves while $E_y$ remains negative. The dashed lines are the estimates given by Eqs.~(\ref{Delta_gamma_general}), (\ref{Delta_x}), and (\ref{Delta_y}).}
       \end{center}
\end{figure}

A general expression for the angle $\theta$ in the regime where the acceleration differs from the purely vacuum case either due to a uniform magnetic field or due to the superluminosity follows from Eq.~(\ref{EQ-balance2}) where we must set $\beta \approx 1$, so that
\begin{equation} \label{EQ-theta-main}
   \theta_* \approx \left[ 2\frac{\delta u}{u} +  \frac{2}{a_0} \frac{\omega_{ce}}{\omega} \right]^{1/2} = \sqrt{\theta_{ph}^2 + \theta_{mag}^2}.
\end{equation}
The applicability condition for this expression is
\begin{equation}
    \theta_* \gg \theta_{vac} \propto \gamma^{-1/2}.
\end{equation}
This condition indicates that the electron would tend to transition into the considered regime with the energy increase. The magnetic field dominates the electron acceleration over the superluminosity of the wave caused by the plasma if $\theta_{mag} \gg \theta_{ph}$, which is equivalent to a requirement that
\begin{equation} \label{dom_B_field}
    \frac{\omega_{ce}}{a_0 \omega} \gg \frac{v_{ph} - c}{c}.
\end{equation}

We are now well-positioned to estimate the energy gain by the electron using Eq.~(\ref{main_2}). It is convenient to re-write this equation as
\begin{equation}
    \frac{d \gamma}{d s} = - \left[ \frac{1}{\omega} \frac{ds}{dt} \right]^{-1} \frac{a_0 p}{\gamma m_e c} \sin \theta \cos \left(s + 2 \pi \psi \right),
\end{equation}
where
\begin{equation}  \label{dephasing_s}
    \frac{1}{\omega} \frac{ds}{dt} = 1 - \frac{1}{v_{ph}} \frac{dx}{dt} = 1 - \frac{\beta}{u} \cos \theta.
\end{equation}
The electron is ultra-relativistic when it starts gaining energy, so that $p/\gamma m_e c \approx 1$. We also use the definitions for $\beta$ and $u$ to obtain that
\begin{equation} \label{dgamma-1}
    \frac{d \gamma}{d s} = - \frac{a_0 u \sin \theta}{u - \beta \cos \theta }  \cos \left(s + 2 \pi \psi \right).
\end{equation}
We assume that the electron starts its acceleration at $E_{wave} < 0$, similarly to what is shown in Fig.~\ref{fig:traj}. This is equivalent to $\cos \left(s + 2 \pi \psi \right) < 0$ at the start of the acceleration and the energy gain continues while this function remains negative. Then the maximum energy gain is estimated by integrating Eq.~(\ref{dgamma-1}) over a phase interval $\Delta s = \pi$ where $\cos \left(s + 2 \pi \psi \right)$ decreases from 0 to -1 and then increases back to 0. We also set $\theta = \theta_*$ to find that
\begin{eqnarray}
    \Delta \gamma &\approx& \frac{2a_0 u \sin \theta_*}{u - \beta \cos \theta_* }.
\end{eqnarray}
We can further simplify this expression by setting $\beta \approx 1$ and taking into account that $\theta_* \ll 1$ and that $u - 1 \ll 1$, which yields
\begin{eqnarray} \label{Delta_gamma_general}
    \Delta \gamma &\approx& \frac{4 a_0 \theta_*}{\theta_*^2 + 2 \delta u}.
\end{eqnarray}
In the regime where the magnetic field determines the electron dynamics [see Eq.~(\ref{dom_B_field})], we have
\begin{eqnarray} \label{Delta_gamma}
    \Delta \gamma_{mag} &\approx& \frac{4 a_0}{\theta_{mag}} = (2 a_0)^{3/2} \left( \frac{\omega}{\omega_{ce}} \right)^{1/2}.
\end{eqnarray}

It is important to point out that a very strong magnetic field reduces the electron energy gain. Indeed, in the regime where the energy gain is determined primarily by the superluminosity, we have 
\begin{eqnarray} \label{Delta_gamma_ph}
    \Delta \gamma_{ph} &\approx& \frac{4 a_0 \theta_{ph}}{4 \delta u} = \frac{2a_0}{\theta_{ph}}.
\end{eqnarray}
Equations~(\ref{Delta_gamma_ph}) and (\ref{Delta_gamma}) can be generalized as $\Delta \gamma \propto 1/\theta$, where $\theta = \max( \theta_{mag}, \theta_{ph})$. This result confirms that, as the magnetic field is increased for a fixed value of $u$ and $\theta_{mag}$ exceeds $\theta_{ph}$, the energy gain becomes dependent on the magnetic field as $B_*^{-1/2}$.

The distance that the electron has to travel with the laser is estimated by estimating the corresponding time interval $\Delta t$ from Eq.~(\ref{dephasing_s}) by setting $ds = \pi$,
\begin{equation}
    \Delta t = \frac{\pi}{\omega} \frac{1}{\delta u + \theta^2/2}.
\end{equation}
We then take into account that $\theta \ll 1$, so that $v_x \approx c$, and find that
\begin{equation}
    \Delta x \approx c \Delta t \approx  \frac{\lambda}{\theta_*^2 + 2 \delta u}.
\end{equation}
An alternative expression in terms of $\Delta \gamma$ from Eq.~(\ref{Delta_gamma_general}) reads
\begin{equation} \label{Delta_x}
    \Delta x / \lambda \approx \Delta \gamma / 4 a_0 \theta_* .
\end{equation}
We estimate the corresponding transverse displacement as $\Delta y \approx \Delta x \tan \theta_*$, which yields
\begin{equation} \label{Delta_y}
    \Delta y /\lambda \approx \frac{\theta_*}{\theta_*^2 + 2 \delta u}.
\end{equation}

We conclude this section by comparing these estimates with the exact solution shown in Fig.~\ref{fig:traj}. Equations~(\ref{Delta_gamma}), (\ref{Delta_x}), (\ref{Delta_y}) for $a_0 = 50$, $\omega_{ce}/\omega = 2.085$, and $\delta u = 0$ yield $\Delta \gamma \approx 690$, $\Delta x / \lambda \approx 12$, and $\Delta y / \lambda \approx 3.5$. These estimates reproduce the dynamics of the electron after it begins to move upwards remarkably well.

%*******************************************************

\section{Parameter scans} \label{Sec-parameter_scan}

In this section we perform parameter scans to determine the predictive capability of the estimates from Sec.~\ref{Sec-estimates}.

\begin{figure}
    \begin{center}
\includegraphics[width=1\columnwidth]{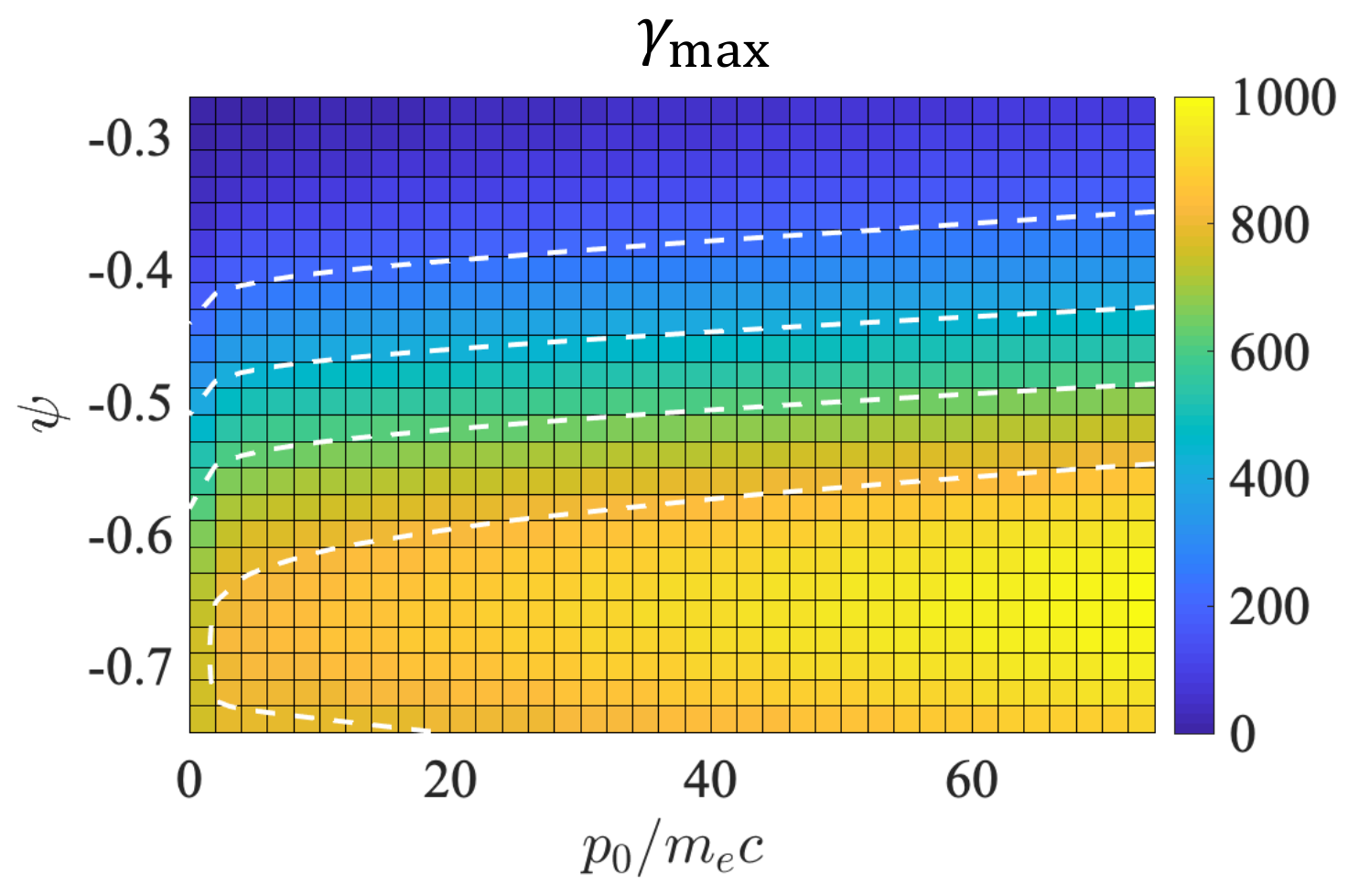}
       \caption{\label{fig:scan_psi_p0} Scan over the phase offset $\psi$ and initial longitudinal momentum $p_0$ for an electron irradiated by a wave with $a_0 = 50$ ($\delta u = 0$) in a magnetic field with $\omega_{ce}/ \omega = 1$. The color shows the maximum relativistic factor $\gamma_{\max}$ that the electron achieves during acceleration (while $E_y$ remains negative). The dashed curves show $\gamma_{max} = 200$, 400, 600, and 800.}
       \end{center}
\end{figure}

Our first scan is over the phase offset $\psi$. We are considering an ultra-relativistic electron with an initial longitudinal momentum $p_0 = 75 m_e c$. The electron begins its motion in a negative electric field of the laser, $E_y(0) < 0$, which implies that $-0.75 < \psi < -0.25$ in Eq.~(\ref{E_wave}). We set $a_0 = 50$ and $\omega_{ce} / \omega = 1$. The electron in this setup starts moving upwards along the $y$-axis and gaining energy, because $(\bm{v} \cdot \bm{E}_{wave}) < 0$. The energy gain continues for as long as $E_y$ remains negative. This agrees with the assumptions that went into our estimates. In order to find the maximum energy gain, we have numerically integrated the equations of motion (\ref{EQ_1}) and (\ref{EQ_2}) for different value of $\psi$ between -0.75 and -0.25. The integration is performed until the electric field becomes positive. The corresponding initial values of $E_y$ and the resulting $\gamma_{\max}$ are shown in Fig.~\ref{fig:phase_scan}. The middle panel of Fig.~\ref{fig:phase_scan} shows the maximum transverse and longitudinal displacements by the electron during the energy gain. The dashed lines are the values given by our estimates [Eqs.~(\ref{Delta_gamma}), (\ref{Delta_x}), and (\ref{Delta_y})]. We conclude that these estimates capture the electron dynamics relatively well, provided that the electron samples a considerable part of the laser cycle with the negative electric field. 

Our second scan whose result is shown in Fig.~\ref{fig:scan_psi_p0} explores the sensitivity to the initial longitudinal momentum $p_0$ in the same setup as in the previous scan and for the same values of $a_0 = 50$ and $\omega_{ce} / \omega = 1$. The key feature here is that the highest value of $\gamma_{\max}$ as a function of $p_0$ remains relatively flat for $p_0 \gg m_e c$. As $p_0$ changes from 5 to 75, the highest value of $\gamma_{\max}$ increases by less than 20\%. The weak dependence that does exist is due to the difference it time that it takes for the electron to reach the regime described by our estimates. A similar trend is observed for the maximum transverse and longitudinal displacements.

\begin{figure}
    \begin{center}
\includegraphics[width=0.9\columnwidth]{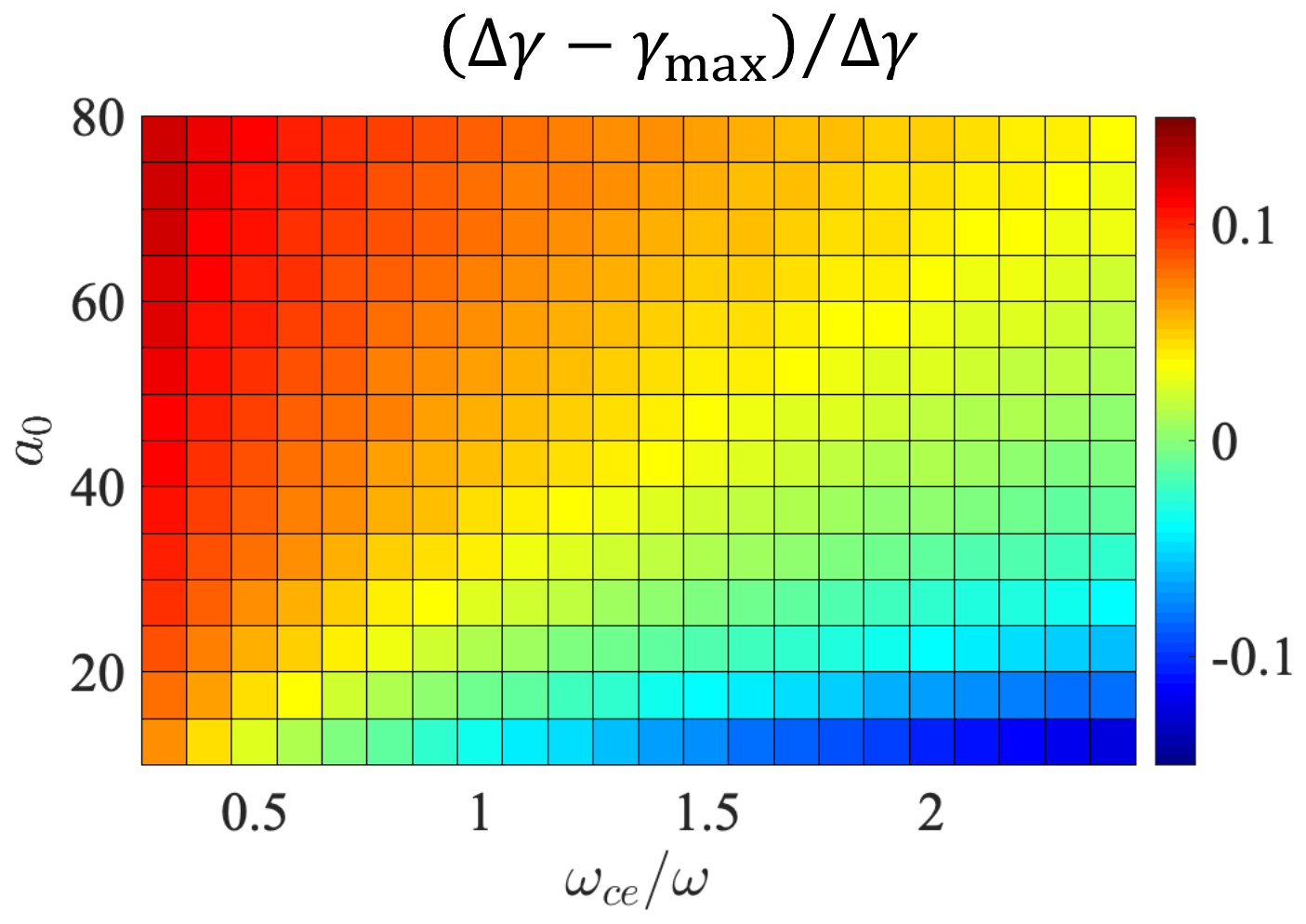}
       \caption{\label{fig:error} Scan over $a_0$ and $\omega_{ce}/ \omega$ at $\psi = -0.65$ and $\delta u = 0$. The initial longitudinal momentum is set at $p_0 = a_0 m_ec$. The color shows a relative difference between the calculated maximum relativistic factor $\gamma_{\max}$ that the electron achieves during acceleration and $\Delta \gamma$ predicted by Eq.~(\ref{Delta_gamma_general}).}
       \end{center}
\end{figure}

Our last scan is over $a_0$ and $\omega_{ce}$ to confirm the derived scaling for the energy gain $\Delta \gamma$ given by Eq.~(\ref{Delta_gamma}). In this case, we fix the phase offset and the ratio between the initial longitudinal momentum $p_0$ and $a_0$ by setting $\psi = -0.65$ and $p_0 = a_0 m_e c$. We find that both the trend and the values predicted by Eq.~(\ref{Delta_gamma}) are reproduced relatively well as we vary $a_0$ from 10 to 80 and $\omega_{ce}/\omega$ from 0.25 to 2.5. Figure~\ref{fig:error} shows the relative error between what we get from the exact solution and what is predicted by Eq.~(\ref{Delta_gamma}). Even though the value of $\gamma_{\max}$ changes by almost two orders of magnitude, the relative error remains below 15\%.

%+++++++++++++++++++++++++++++++++++++++++++++++++++++++++++++++++++++

\section{Impact of superluminosity} \label{Sec-superluminosity}

The estimates provided in Sec.~\ref{Sec-estimates} include not only the magnetic field but also the phase velocity $v_{ph}$ because they both have a similar impact on the electron acceleration.

\begin{figure}
    \begin{center}
\includegraphics[width=0.99\columnwidth]{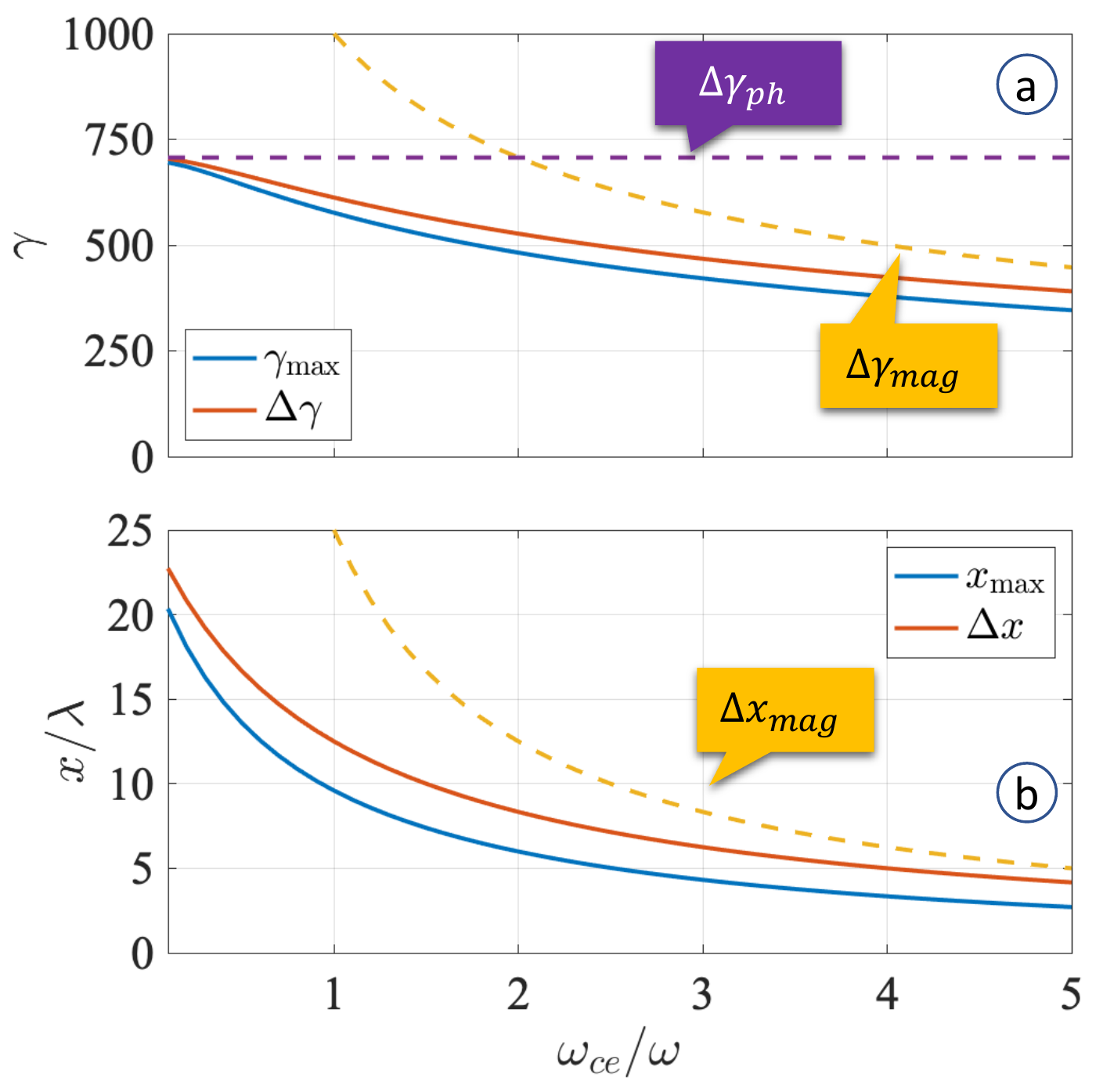}
       \caption{\label{fig:7} Scan over the magnetic field strength for an electron irradiated by a wave with $a_0 = 50$ and $\delta u = 0.01$. The phase offset is $\psi = -0.7$ and the initial longitudinal momentum is $p_0 = 10 m_ec$. The red curves are the calculated $\gamma_{\max}$ and the corresponding longitudinal displacement $x_{\max}$. The blue curves are $\Delta \gamma$ and $\Delta x$ predicted by Eqs.~(\ref{Delta_gamma_general}) and (\ref{Delta_x}). $\Delta \gamma_{mag}$ and $\Delta \gamma_{ph}$ are given by Eqs.~(\ref{Delta_gamma}) and (\ref{Delta_gamma_ph}). $\Delta x_{mag}$ is given by Eq.~(\ref{Delta_x}) with $\theta = \theta_{mag}$.}
       \end{center}
\end{figure}

In order to illustrate more clearly the impact of the superluminosity, we have performed a scan over the strength of the magnetic field for a fixed value of $\delta u = 10^{-2}$. In this case, $a_0 = 50$ and the initial longitudinal momentum is set to $p_0 = 10 m_ec$. The phase offset is also fixed at $\psi = -0.7$. The result is shown in Fig.~\ref{fig:7}, where we show how the maximum $\gamma$-factor and the longitudinal displacement during the acceleration change with $\omega_{ce}/\omega$. Note that we again run the calculation only while $E_y$ is negative, because the change in sign of $E_y$ terminates the electron acceleration in this example. The red curves in Fig.~\ref{fig:7} are $\Delta \gamma$ and $\Delta x$ predicted by Eqs.~(\ref{Delta_gamma_general}) and (\ref{Delta_x}). These estimates agree relatively well with the result of the exact calculation. 

The dashed curves in Fig.~\ref{fig:7}a represent two limiting regimes: the regime where the energy gain is limited by the magnetic field (yellow) and the regime where the energy gain is limited by the superluminosity (purple). The two curves intersect at
\begin{equation} \label{main_condition}
    \omega_{ce} / \omega = 4 a_0 \delta u.
\end{equation}
At $\omega_{ce} / \omega \gg 4 a_0 \delta u$, the magnetic field is sufficiently strong to negate the effect of the superluminosity and one can set $\delta u = 0$ to simplify the analysis. This condition is consistent with that given by Eq.~(\ref{dom_B_field}). However, the superluminosity significantly limits the energy gain by the electron at $\omega_{ce} / \omega \leq 4 a_0 \delta u$ and must be taken into account. As seen from Fig.~\ref{fig:7}a, the upper limit on the energy gain for a fixed value of $\delta u$ is given by $\Delta \gamma_{ph}$ from Eq.~(\ref{Delta_gamma_ph}). Figure~\ref{fig:7}b shows that the reduction in the energy gain is associated with s reduction of the distance travelled by the electron before reaching the maximum energy gain. It is significantly shorter than $\Delta x_{mag}$ given by Eq.~(\ref{Delta_x}) that assumes $\delta u = 0$.

\begin{figure}
    \begin{center}
\includegraphics[width=1\columnwidth]{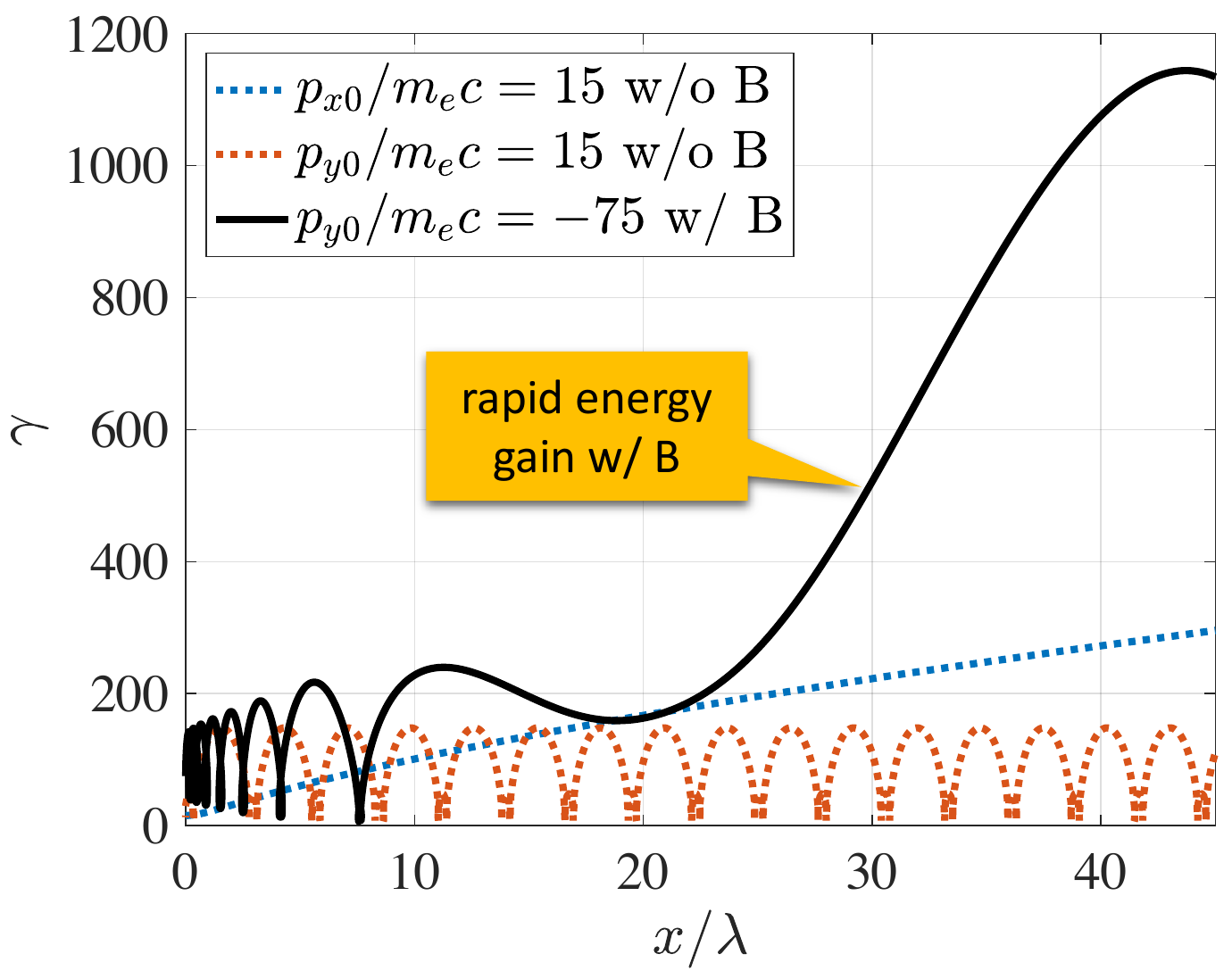}
       \caption{\label{fig:10} Electron energy gain with and without the applied magnetic field. In all three cases we have $a_0 = 50$ and the electron starts its motion at $E_y = E_0$. The blue and red dotted curves are for the acceleration without the magnetic field with an initial longitudinal and an initial transverse momentum, respectively. The solid curve is for an electron with an initial transverse momentum $p_0 = - 75 m_e c$ accelerated in a magnetic field, $\omega_{ce}/\omega = 1.01$.}
       \end{center}
\end{figure}

One source of the superluminosity is the presence of the plasma itself. In a cold plasma, a linear plane electromagentic wave has the following dispersion relation:
\begin{equation}
    \omega^2 = \omega_{pe}^2 + k^2 c^2,
\end{equation}
where $k$ is amplitude of the wave-vector and $\omega_{pe} = \sqrt{4 \pi n_e e^2 / m_e}$ is the plasma frequency for electrons with density $n_e$. In the limit of $v_{ph} - c \ll c$, we have
\begin{equation} \label{superlum}
    \delta u = \frac{v_{ph} - c}{c} \approx \frac{1}{2}\frac{\omega_{pe}^2}{\omega^2} = \frac{1}{2}\frac{n_e}{n_{crit}},
\end{equation}
where $n_{crit}$ is the cutoff electron density (often called the critical density) determined by the condition $\omega_{pe} = \omega$. A laser pulse of relativistic intensity can make a plasma relativistically transparent by heating the electrons to relativistic energies. This aspect can be taken into account by adjusting Eq.~(\ref{superlum}), with the superluminosity related to the relativistic transparenty given by
\begin{equation} \label{superlum_RT}
    \delta u_{RT} \approx \frac{1}{2a_0}\frac{\omega_{pe}^2}{\omega^2} = \frac{1}{2a_0}\frac{n_e}{n_{crit}}.
\end{equation}
The relation given by Eq.~(\ref{main_condition}) now reads
\begin{equation} \label{main_condition2}
    \omega_{ce} / \omega = 4 a_0 \delta u_{RT} \approx 2 \frac{n_e}{n_{crit}}.
\end{equation}
We therefore conclude that a static magnetic field determines the electron energy gain in a relativistically transparent plasma if its strength satisfies the condition
\begin{equation} \label{main_condition2_RT}
    \omega_{ce} / \omega \gg 2 n_e/n_{crit}.
\end{equation}
In the case of a 1~$\mu$m laser, this condition can be re-written as
\begin{equation}
    B_* [\mbox{kT}] \gg 20 n_e/n_{crit}.
\end{equation}

%+++++++++++++++++++++++++++++++++++++++++++++++++++++++++++++++++++++

\section{Summary and discussion} \label{Sec-summary}

\begin{figure}
    \begin{center}
\includegraphics[width=1\columnwidth]{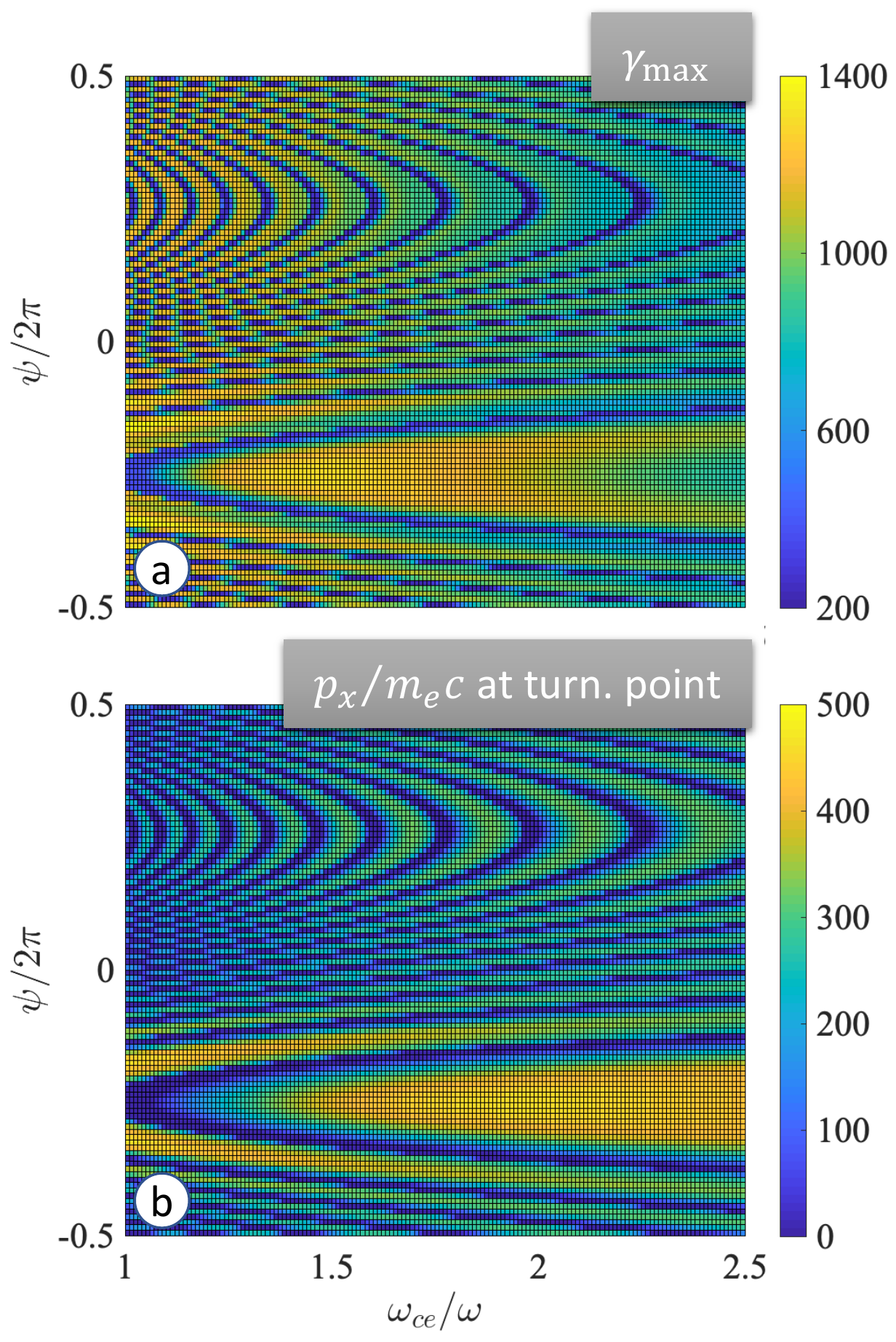}
       \caption{\label{fig:8} Parameter scan for a laser-irradiated electron that starts its motion with a transverse momentum. The laser amplitude is $a_0 = 50$ and the initial momentum is $p_0 / m_ec = 75 - a_0 \sin(\psi)$.}
       \end{center}
\end{figure}

We have considered electron acceleration by a laser pulse in a plasma with a static uniform magnetic field $B_*$. In our setup, the laser pulse propagates perpendicular to the magnetic field lines with the polarization chosen such that $(\bm{E}_{laser} \cdot \bm{B}_*) = 0$. The focus of the work is on electrons with an appreciable initial transverse momentum, $p_0 \sim a_0 m_e c$. These electrons are unable to gain significant energy from the laser pulse in the absence of the magnetic field due to strong dephasing (see the red dotted curve in Fig.~\ref{fig:10}). We have shown that the magnetic field can initiate an energy increase by rotating the electron, such that its momentum becomes directed forward. 

We found that the energy gain continues well beyond the turning point where the dephasing drops to a very small value due to the momentum rotation induced by the magnetic field. In contrast to the case of purely vacuum acceleration, the electron continues to move at a significant angle with respect to the laser propagation as its energy increases. The maximum energy gain given by Eq.~(\ref{Delta_gamma_general}) depends not only on the strength of the magnetic field but also on the phase velocity of the wave. The magnetic field is the limiting factor if its strength exceeds the value given by Eq.~(\ref{main_condition}). Otherwise, the energy gain is limited by the superluminosity.

A distinctive feature of the discussed electron acceleration mechanism is a rapid energy gain compared to what is possible with pure vacuum acceleration. Figure~\ref{fig:10} shows the relativistic factor $\gamma$ as a function of the longitudinal coordinate for electrons that are accelerated at $a_0 = 50$ with and without the applied magnetic field. In the absence of the magnetic field, the electron with $p_{x0} = 15 m_ec$ has a reduced dephasing rate and is able to experience a prolonged acceleration. The reduced dephasing is similar to what happens in the magnetic field at the turning point. However, the energy gain is relatively slow, as the electron reaches only $\gamma \approx 300$ after traveling 45 $\mu$m with the laser pulse. In contrast to that, the electron accelerated in the magnetic field experiences a rapid energy gain after the turning point, with an increase of $\Delta \gamma \approx 1000$ over just 20~$\mu$m. It is worth pointing out that, in principle, the energy gain can be very rapid during the vacuum acceleration, but this comes at the expense of the maximum energy gain. The red dotted curve in Fig.~\ref{fig:10} illustrates this for an electron that has an initial transverse momentum.

\begin{figure}
    \begin{center}
\includegraphics[width=1\columnwidth]{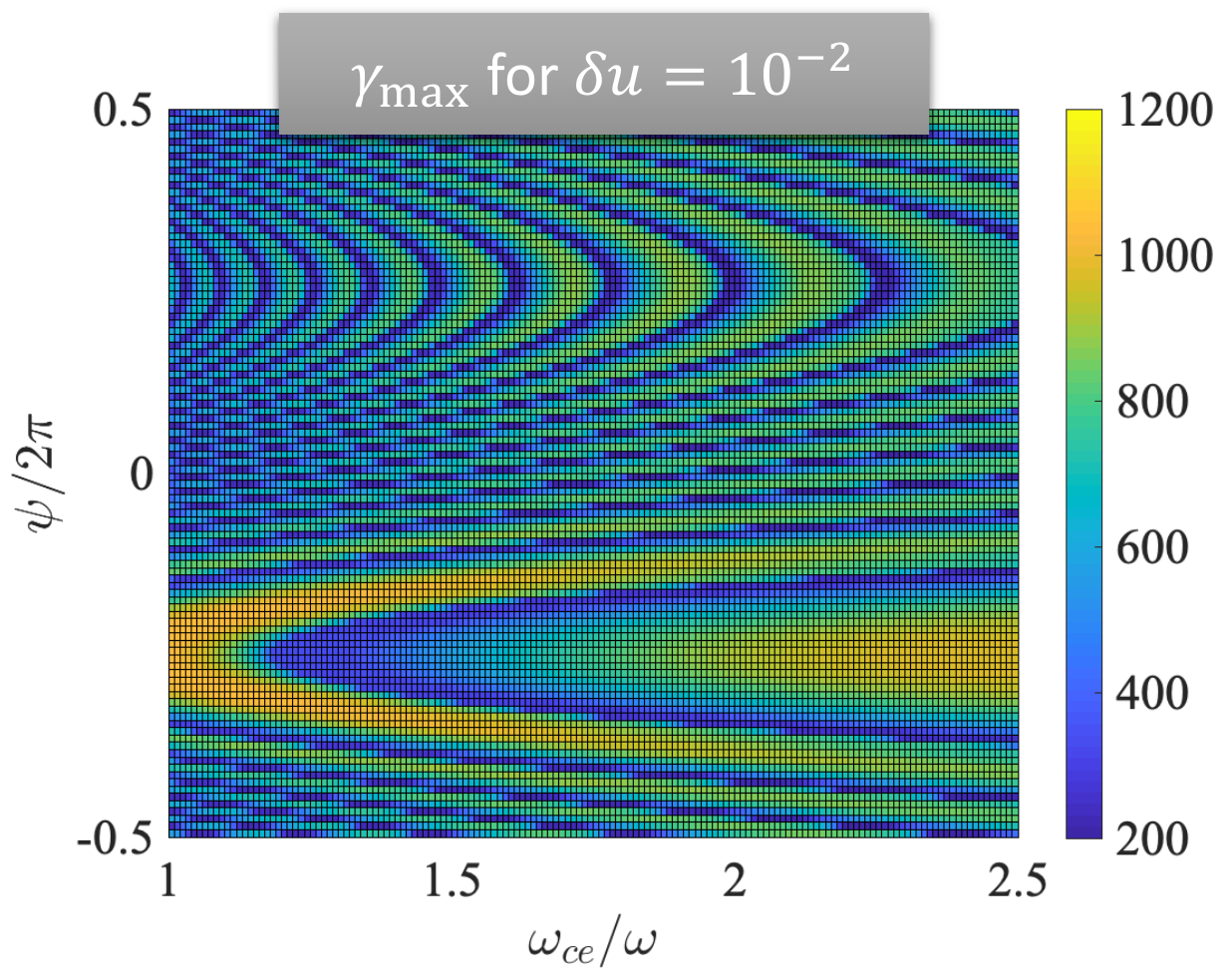}
       \caption{\label{fig:9} Parameter scan for an electron irradiated by a superluminal electromagnetic wave with $\delta u = 0.01$ and $a_0 = 50$. The electron starts its motion with a transverse momentum $p_0 / m_ec = 75 - a_0 \sin(\psi)$. The color shows the maximum relativistic factor along the trajectory similar to that shown in Fig.~\ref{fig:traj}.}
       \end{center}
\end{figure}

The energy enhancement by the magnetic field can be particularly useful at high laser amplitudes, $a_0 \gg 1$, where the acceleration similar to that in the vacuum is unable to produce energetic electrons over tens of microns. A strong magnetic field can help leverage an increase in the laser intensity without a significant increase in the interaction length. The results reported in Ref.~[\onlinecite{Willingale_2018}] for generation of energetic electrons in near-critical plasmas is a relevant example of a strong magnetic field enhancing electron energy gain over a relatively short distance. In this paper, we have considered only one passage by the electron through a turning point and the subsequent acceleration. Multiple passes can further increase the energy gain, but the heating can become stochastic. 

It is important to stress that the energy gain is conditional on the electron having a relativistic longitudinal momentum at the turning point. Figure~\ref{fig:8} shows a parameter scan over the initial phase offset $\psi$ and $\omega_{ce}/\omega$. The modulations of $\gamma_{\max}$ in Figure~\ref{fig:8}a are directly correlated with the changes in $p_x/m_e c$ at the turning point shown in Figure~\ref{fig:8}b. As shown in Fig.~\ref{fig:traj_p}, the longitudinal momentum of the electron is modulated by the laser, so the travel time to the turning point determines the corresponding $p_x/m_e c$. A similar pattern is observed in the case with a superluminal wave shown in Figure~\ref{fig:9}. The implication of this observation is that the energy gain can be suppressed compared to what is predicted by Eq.~(\ref{Delta_gamma_general}) if the electron arrives at the turning point with a low longitudinal momentum. Therefore, our result provides and upper estimate without accounting for the global electron dynamics prior to the onset of the acceleration.

%*******************************************************

\section*{Acknowledgements}
This research was supported by the DOE Office of Science under Grant No. DE-SC0018312. Z.G. was supported in part by the scholarship from China Scholarship Council (CSC) under Grant CSC No. 201706010038

\section*{References}
\bibliography{main}
%\bibliography{Collection}
%###########################################################################################

\end{document}